\documentclass[twocolumn,twoside,slac_two]{revtex4}
\usepackage{graphicx}
\usepackage{fancyhdr}
\begin{document}

\title{Optimization of Integrated Luminosity of the Fermilab Tevatron Collider}

%

\author{M.E. Convery}
\affiliation{Fermi National Accelerator Laboratory, Batavia, IL 60510, USA}

\begin{abstract}
We present the strategy which has been used recently to optimize
integrated luminosity at the Fermilab Tevatron proton-antiproton
collider. We use a relatively simple model where we keep the proton
intensity fixed, use parameters from fits to the luminosity decay of
recent stores as a function of initial antiproton intensity (stash
size), and vary the stash size to optimize the integrated luminosity
per week. The model assumes a fixed rate of antiproton production, that
a store is terminated as soon as the target stash size for the next
store is reached, and that the only downtime is due to store
turn-around time. An optimal range of stash sizes is predicted. Since
the start of Tevatron operations based on this procedure, we have seen
an improvement of approximately 35\% in integrated luminosity.  Other
recent operational improvements have been achieved by decreasing the
shot-setup time and by reducing beam-beam effects by making the proton
and antiproton brightnesses more compatible, for example by scraping
protons to smaller emittances.
\end{abstract}

\maketitle

\thispagestyle{fancy}


\section{Introduction}

\begin{figure}[h]
\includegraphics[width=80mm]{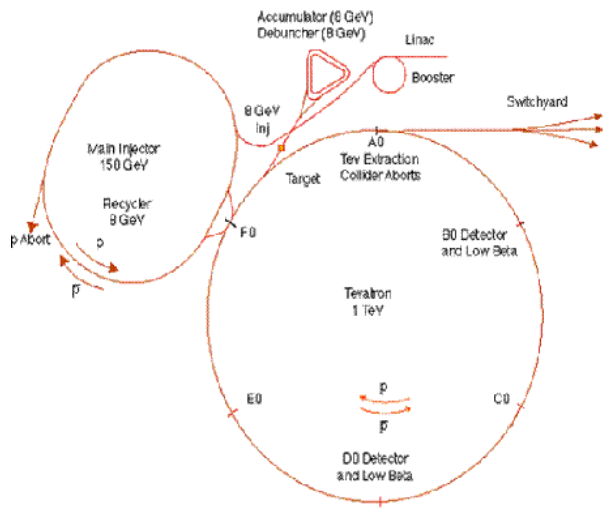}
\caption{The Fermilab accelerator complex.} \label{acceleratorcomplex}
\end{figure}

The Fermilab accelerator complex (Fig.~\ref{acceleratorcomplex}) provides beam to two collider experiments, CDF and D0 at the Tevatron, two neutrino experiments, MiniBooNE and NuMI, and 120 GeV fixed-target experiments.  

The proton source consists of a Cockroft-Walton which accelerates H$^-$ ions to 750 keV, Linac which accelerates the ions to 400 MeV, and the Booster ring.  Multiple turns worth of beam are loaded into the Booster, with the electrons being stripped off the ions on the first turn, and then the accumulated proton beam being accelerated to 8 GeV.  MiniBooNE receives 8 GeV protons on target from the Booster.  The protons otherwise continue to the Main Injector, where they can be accelerated to 120 GeV and sent to the Pbar Target for antiproton production, to NuMI, or through the Transfer Hall in the Tevatron ring to the Switchyard fixed-target area.  Protons for the collider are accelerated to 150 GeV in the Main Injector and then transferred to the Tevatron, where, once antiprotons are also loaded, they are accelerated to 980 GeV. 

Antiprotons are produced by sending 120 GeV protons into a target.  The particles produced are focused using a lithium lens, separated by charge using a bend magnet, and then negatively-charged particles are sent down a transport line, at the end of which only antiprotons remain, to the antiproton Debuncher and Accumulator.  Antiprotons are stored on a short timescale in the Accumulator, where they are cooled stochastically, and then are transferred to the Recycler storage ring in the same tunnel as the Main injector, where they are stored until enough antiprotons have been produced for a new store of protons and antiprotons in the Tevatron.  One distinct advantage of the Recycler is electron cooling, where a beam of electrons passes along side of the antiproton beam in a portion of the ring.  Coulomb scattering between the antiprotons and electrons brings the antiprotons into thermal equilibrium with the electrons, and by continually refreshing the electron beam, brings the antiproton momenta into a tighter range around the desired momentum.  Besides this advantage for storing antiprotons in the Recycler, antiprotons are also produced much more effectively when only a small number is present in the Accumulator.  Antiprotons are injected into the Tevatron by transferring the 8 GeV antiprotons from the Recycler to the Main Injector, accelerating them to 150 GeV, and then transferring them to the Tevatron, 4 bunches at a time.

Antiprotons in the Accumulator are referred to as the {\bf stack}, and {\bf stacking} is the term we use for producing them.  A {\bf pbar} (${\bar p}$) {\bf transfer} is the process of moving antiprotons from the Accumulator to the Recycler, and once they are in the Recycler, they are referred to as the {\bf stash}.  A {\bf store} is a colliding set of protons and antiprotons in the Tevatron (36 bunches of each), and {\bf collider shot setup} is the process of loading a store into the Tevatron.

The luminosity is a measure of the number of collisions expected.  For an intersecting storage ring collider, the instantaneous luminosity is given by ${\cal L}=fnN_1N_2/A$, where $f$ is the revolution frequency of a bunch, $n$ is the number of bunches in each beam, $N_{1(2)}$ is the number of particles in each bunch [protons(antiprotons) in the case of the Tevatron], and $A$ is the cross section of the beam.  The instantaneous luminosity decays over time as the number of particles decreases due to collisions and other losses and the size of the beam in phase space increases.  At the Tevatron, the instantaneous luminosity is in the range of $10^{32}$ cm$^{-2}$ s$^{-1}$, and the integrated luminosity over the course of a store is on the order of a few pb$^{-1}$.

\section{Model for Optimization of Integrated Luminosity \label{model}}

At this point in the collider Run II, all major upgrades have been incorporated and the antiproton stacking rate is not expected to undergo any more large increases.  With relatively stable and reproducible conditions, we ask ourselves how to make the most of what we have.  We take the antiproton production rate to be the limiting factor in integrated luminosity, and by using recent historical data to model the performance of the accelerator complex, find the optimal use of antiprotons for maximizing integrated luminosity.

In the model, proton parameters are kept fixed since they have little variation in practice, with intensities around $320\times 10^9$ per bunch and emittances (a measure of their area in phase space) around $16-17 \pi$ mm mrad at 8 GeV in the Main Injector.  

Luminosity parameters are obtained using data from recent stores.  The dependence of initial luminosity on the number of antiprotons in the stash is quite reproducible, as is the luminosity lifetime behavior, which also happens to be roughly independent of initial luminosity.  

For the antiprotons, the model takes into account the effective production rate, including the stacking rate, the pbar transfer efficiency, lifetimes in both the Accumulator and Recycler, and any interruption to stacking during pbar transfers, and also the efficiency of antiproton transfers to the Tevatron.  The model calculates antiproton production and integrated luminosity over the course of a week given a target stash size at which the existing store is terminated and the antiprotons are transferred to the Tevatron for a new store.  

\begin{figure}[h]
\includegraphics[width=\linewidth]{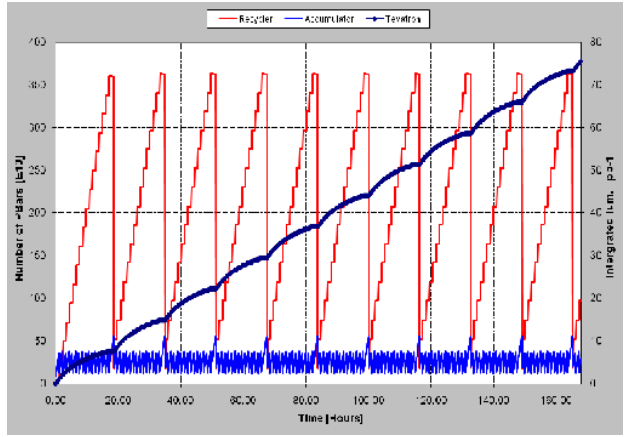}
\caption{Number of antiprotons in the Accumulator (blue) and Recycler (red) along with integrated luminosity (navy) in a modeled week.} \label{modelweek}
\end{figure}

\begin{figure}[h]
\includegraphics[width=\linewidth]{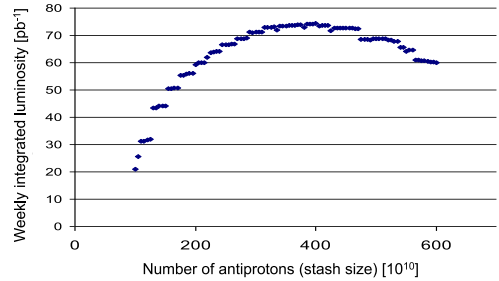}
\vglue -0.9in 
\caption{Output of the model showing predicted weekly integrated luminosity as a function of target stash size.} \label{modelresult}
\end{figure}

\begin{figure}[htb]
\includegraphics[width=80mm]{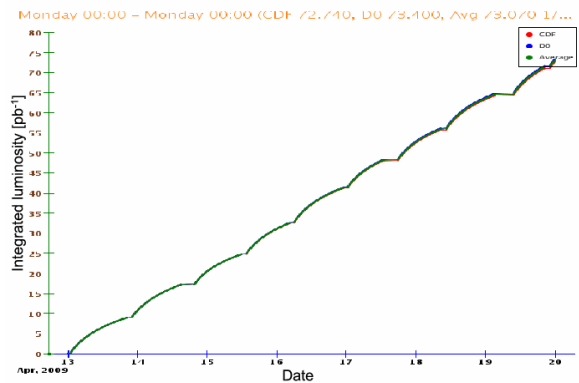}
\vglue -0.3in 
\caption{Luminosity integrated over the week of April 13, 2009, which totaled 73 pb$^{-1}$.} \label{perfectweek}
\end{figure}

\begin{figure*}[htb]
\includegraphics[width=0.9\linewidth]{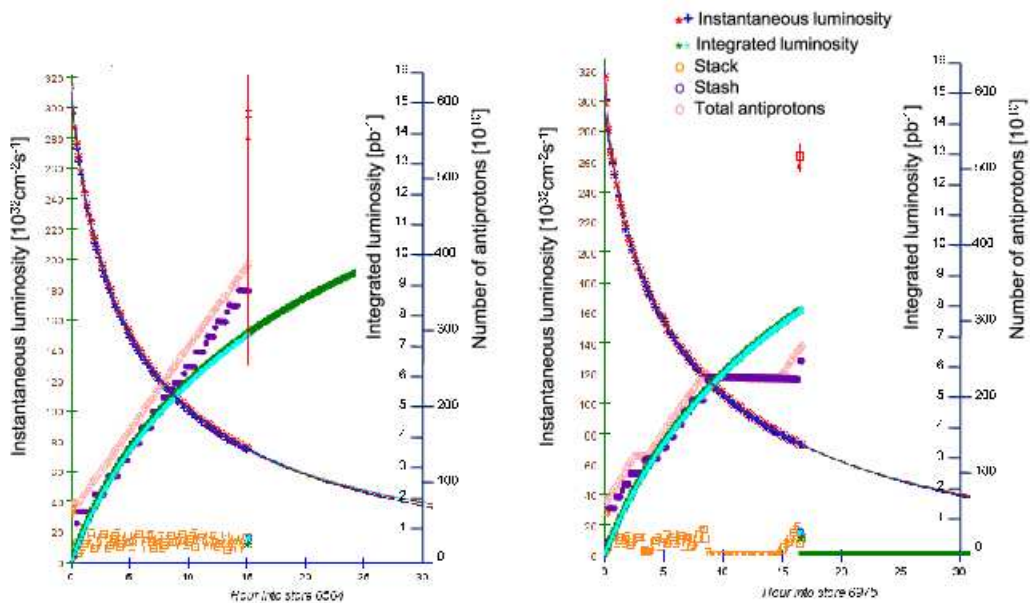}
\vglue -1.4in
\caption{Output of a tool used to determine store duration showing instantaneous and integrated luminosity over the course of a collider store along with the number of antiprotons available, which predicts future integrated luminosity of the current store and of a potential new store from the antiprotons available.} \label{lumdecay}
\end{figure*}

Figure~\ref{modelweek} shows the number of antiprotons in the stack and stash as well as the integrated luminosity over the course of a week for a target stash size of $\sim 360\times 10^{10}$, a peak stacking rate of $30\times 10^{10}/{\mbox hr}$, and a collider shot setup time of 1.5 hrs.  Figure~\ref{modelresult} shows the output weekly integrated luminosity as a function of input target stash size.  The model indicates that the integrated luminosity is maximized by a target stash size of $\sim 350-400 \times 10^{10}$.  These are the conditions under which the accelerator complex operated for the last year or so.  The predicted maximum weekly integrated luminosity of $\sim 74$ pb$^{-1}$ was found to be quite accurate when compared with two ``perfect'' weeks in which we had very little downtime; the luminosity integrated in one of those weeks, totaling 73 pb$^{-1}$, is shown in Fig.~\ref{perfectweek}.

An additional tool is available which is used for store-by-store decision making and which also confirms the conclusion drawn from the model discussed above.  Figure~\ref{lumdecay} shows the output of that tool, which is updated hourly during the course of a store.  Red and blue points show the instantaneous luminosity as measured by CDF and D0, and curves fitting the luminosity decay are also shown.  Green points show the integrated luminosity as determined from the decay fit, and as measured by CDF is shown in cyan.  The number of antiprotons in the stack is shown in orange, in the stash in purple, and total in pink.  The tool predicts the initial luminosity of a new store from the number of antiprotons available based on recent stores from similar stash size and plots that as a separate red point.  The integrated luminosity over the first hour of that potential new store is also predicted based on the same historical data, and is compared to the luminosity which would be integrated over the next 2.5 hours of the current store based on the luminosity decay fit (comparing 2.5 hours of the existing store versus 1.5 hours of shot setup plus 1 hour of the the store).

In the left plot of Fig.~\ref{lumdecay}, the number of antiprotons available is around our typical target stash of $375\times 10^{10}$, the predicted luminosity of the current store over 2.5hrs is about 600 nb$^{-1}$, while the predicted luminosity over the first hour of a new store is about 800 nb$^{-1}$, confirming that when we reach the target stash size determined from our model, putting in a new store will lead to more integrated luminosity than continuing to run the existing one.  The plot on the right of Fig.~\ref{lumdecay} shows that this tool is especially useful during non-standard running conditions, such as when we have long stacking downtimes.  We find there that although the stash is only about $275\times 10^{10}$, the luminosity has decayed enough that we would be integrating more with a new store than with the existing one (790 nb$^{-1}$ vs.\ 750 nb$^{-1}$).

This tool is used to make store-by-store operational decisions, especially in response to interruptions of standard operating conditions, while the model directs our general plan to maximize integrated luminosity over the time scale of multiple stores.  By varying conditions in the model, we also gain insight into areas to attack in order to improve integrated luminosity.  The model is also rerun whenever changes in the performance the accelerator complex occur in order to ensure that our operations are optimized.  The model is described in more detail in Ref.~\cite{cons}.

The model determines the best target stash size given certain conditions, such as antiproton production, shot setup duration, proton intensity and emittances, and the behavior of the Tevatron in initial luminosity vs.\ stash size and luminosity decay.  If we can improve these conditions, we can make additional gains.  Section~\ref{antiproton} describes additional efforts to optimize antiproton production, and Sec.~\ref{otheroperational} describes some recent operational improvements in these other areas.

\begin{figure*}[htb]
\includegraphics[width=\linewidth]{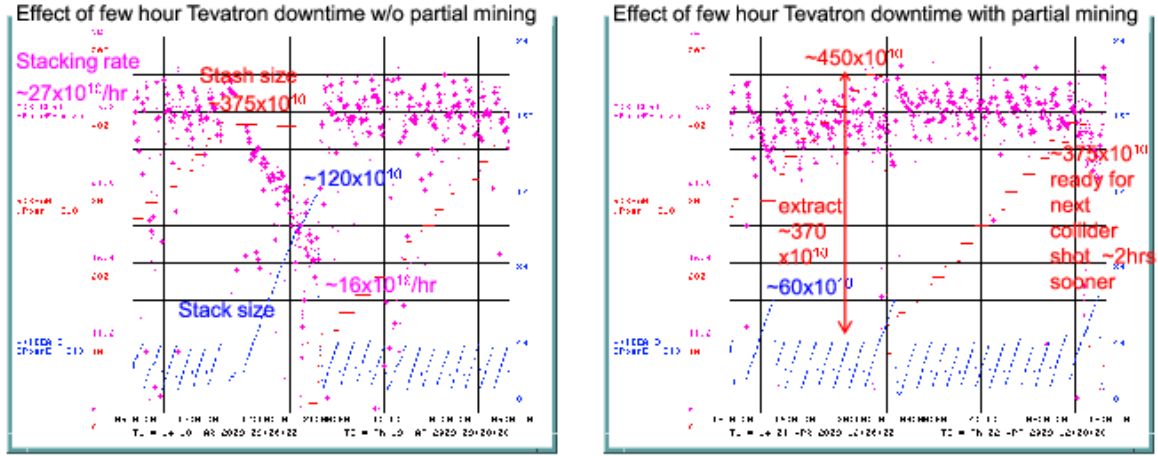}
\vglue -2.5in
\caption{Antiproton stacking rate, stack, and stash size as a function of time over the course of a collider store, a Tevatron down time, and another collider store, with and without ``partial mining'' of antiprotons from the Recycler.} \label{partialmining}
\end{figure*}

\section{Optimization of Antiproton Production \label{antiproton}}

\subsection{Optimizing pbar transfers}

One feature of overall antiproton production over which we have a lot of control is the stack size at which a pbar transfer from Accumulator to Recycler is initiated.  In favor of frequent transfers from small stacks is the fact that the stacking rate declines as the stack size increases.  Up to stack sizes of $\sim 40\times 10^{10}$, the rate is around 25-30$\times 10^{10}/$hr, while at a stack size of $80\times 10^{10}$, the rate drops to about $20\times 10^{10}/$hr \cite{brian}.  However, frequent transfers would not be optimal if there were long durations in which we were not stacking during the transfer process.  Other factors related to the transfer process are the percentage of antiprotons removed from the stack and the transfer efficiency to the Recycler, both of which depend on the stack size and the number of transfers in a set.  The lifetime of antiprotons in the Recycler is also a factor, and we must take into consideration the need for adequate cooling between the last pbar transfer and a collider shot.  Given the current conditions, we found that a set of two transfers initiated when the stack reached $25\times 10^{10}$ was more optimal than our previous mode with a varying number of transfers from a stack of $40\times 10^{10}$.

\subsection{Rapid transfers}

As mentioned, non-stacking time during a pbar transfer makes frequent transfers inefficient.  Much effort has been put into speeding up pbar transfers and improving transfer efficiency over the past several years \cite{jim}.  The time to prepare and execute the transfer has been reduced from as much as an hour down to less than 5 minutes, with the non-stacking time now being negligible.  At the same time, transfer efficiencies have increased from 80-90\% to an average of 95\%.

\subsection{``Partial mining''}
Another gain in overall antiproton production has been achieved though a new operational method in the Recycler referred to as ``partial mining'', which enables only a percentage of the stash to be extracted (``mined'') for a collider shot, without compromise of cooling or lifetime of the antiprotons.  In the new method, RF manipulations separate the beam to extracted from the beam to be left behind in the Recycler.  Because of the RF bucket size, there are limitations such that the fraction of beam extracted must be in the range 20-80\%, and no more than $150\times 10^{10}$ can be left behind.  This capability allows us to maintain regular pbar transfers even after the target stash size has been reached, thus maintaining small stack sizes in the Accumulator and the associated higher stacking rates and better pbar transfer efficiencies.  In addition to the gains in overall antiproton production, this gives us more flexibility in scheduling collider shots to work around problems, and can give us a backup supply of antiprotons if we have a failure during collider shot setup.

Figure~\ref{partialmining} shows the effect of a few hour Tevatron downtime with and without partial mining.  On the left, we see that without partial mining, the target stash size of $375\times 10^{10}$ is reached and instead of continuing pbar transfers, the stack size grows to $120\times 10^{10}$ with the stacking rate dropping from $27\times 10^{10}/$hr down to $16\times 10^{10}/$hr.  Once the Tevatron is ready, the stash is transferred to the Tevatron, the large stack is transferred to the Recycler, and regular stacking and transfers resume.  At the end of the time period shown on the plot, the target stash size for a new collider shot is just reached.  On the right-hand plot, after the target stash is reached, pbar transfers continue as usual and the stacking rate remains high.  When the Tevatron is ready, the stash has reached $450\times 10^{10}$, and partial mining allows us to extract only $370\times 10^{10}$ for the collider shot.  Regular pbar transfers continue, and we have reached the target stash for yet another collider shot about 2hrs sooner than in the previous example, thus trading 2hrs of lower luminosity at the end of a store for the higher luminosity of an new store, leading to more integrated luminosity overall.

\section{Other Operational Improvements \label{otheroperational}}

\subsection{Reducing collider shot setup time}
When we implemented the model described in Sec.~\ref{model}, where the store duration was determined by the target stash size of $\sim 375\times 10^{10}$, the average store duration dropped from $\sim 20$hrs down to 15hrs.  With shorter and more frequent stores, the amount of time spent in collider shot setup becomes more of a factor in overall integrated luminosity.  A task force was created to shorten shot setup time.  The biggest time savings came in the antiproton load, achieved by timeline changes and Recycler RF manipulations, and in the proton load achieved by multi-batch coalescing in the Main Injector, allowing bunches to be injected two at a time.  Other changes in Tevatron tune-up and settings are in progress.  The collider shot setup time has been reduced from approximately 2hrs down to a little over an hour.

In addition to the increase in integrated luminosity by taking less time between stores, shorter shot setups also mean less time spent at reduced rates of protons on target to stacking and to NuMI.  

\subsection{Increasing proton brightness}

Brighter beam means more particles in a smaller area; it is defined as intensity over emittance.  Increasing brightness leads to higher instantaneous luminosity.  In the Tevatron, because of very small antiproton emittances achieved in the Recycler, the antiproton beam is brighter than the proton beam.  A big difference in brightness leads to beam-beam effects, where the bright antiproton beam shifts the proton beam in phase space where it may be less stable.  We intentionally ``blow up'' the antiproton emittances in the Tevatron before collisions in order to better match the proton emittance.  Brighter protons would therefore increase instantaneous luminosity, reduce losses due to beam-beam effects, and allow brighter antiprotons which increases luminosity again.

\begin{figure}[h]
\includegraphics[width=1.8\linewidth]{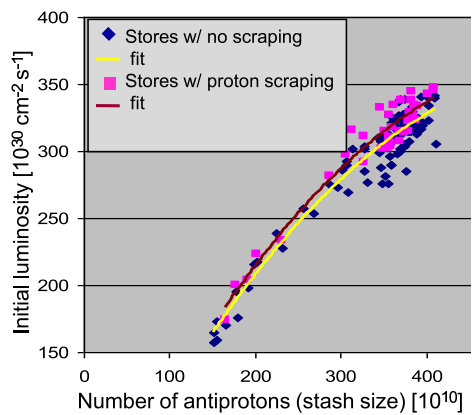}
\vglue -2in
\caption{Effect of proton scraping on initial luminosity as a function of number of antiprotons used for the collider shot.} \label{protonscraping}
\end{figure}

We have achieved brighter protons by scraping the proton halo with collimators in the Main Injector before they are accelerated and injected into the Tevatron.  That is, we start with higher intensity beam, and scrape to nominal intensity but smaller emittance.  As shown in Fig.~\ref{protonscraping}, this has improved initial luminosities by about 3-4\%.  Smaller proton emittances also lead to improved transfer and acceleration efficiencies, and the improved dynamic aperture of the machine has reduced quenching, where beam falls out of the machine catastrophically.

\begin{figure}[h]
\includegraphics[width=\linewidth]{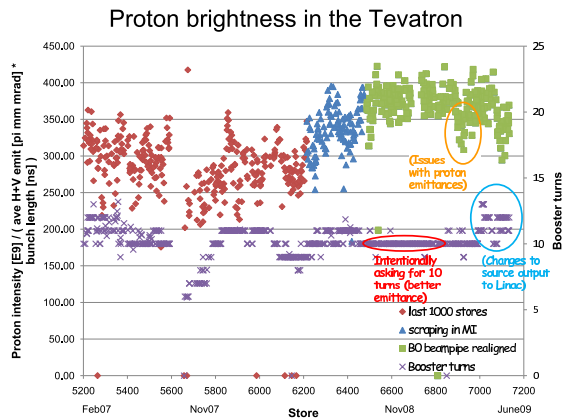}
\caption{Proton brightness in the Tevatron as a function of time, showing the effect of operational and machine improvements.} \label{protonbrightness}
\end{figure}

The proton brightness is shown in Fig.~\ref{protonbrightness} as a function of time (referenced by store number).  Along with scraping protons in the Main Injector, an effect was observed dependent on the number of turns worth of protons loaded into the Booster.  Although 11 turns led to higher intensity than 10, the emittance was also significantly larger, such that the overall brightness was higher for 10 turns.  Also marked is the period following the realignment of a section of beampipe in the CDF collision hall which was an aperture restriction.  This allowed us to go to even higher proton brightness without quenching.  At high antiproton intensities, where beam-beam effects are an issue, the fraction of protons surviving acceleration in the Tevatron went from 95-97\% to 97-98\% with the removal of the aperture restriction.

\subsection{Consistency and reliability}

Following the guidance of the model discussed in Sec.~\ref{model}, we have a target stash size for collider shot setups.  Using that consistent stash size, Recycler cooling and Tevatron tunes are also consistent from store to store.

Tevatron stability has also been improved by automatically setting the proton tune based on antiproton intensity, an orbit stabilization program, and monitoring lattice stability.  Beam-beam effects have been reduced by controlling the antiproton/proton emittance ratio, both by ``blowing up'' antiprotons and scraping protons.  The realignment of the beampipe near CDF to remove the aperture restriction has also improved stability.

\begin{figure}[h]
\includegraphics[width=1.8\linewidth]{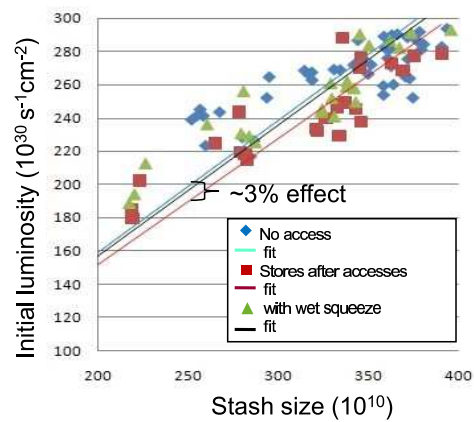}
\vglue -2in
\caption{Effect of a wet squeeze after a collision hall access on initial luminosity as a function of number of antiprotons used for the collider shot.} \label{wetsqueeze}
\end{figure}

Another area where we have improved reliability is recovering from a collision hall access.  After the experiments make an access, there is an overhead of approximately 2hrs before we are ready for collider shot setup.  Since the low-beta quadrupoles inside the collision halls must be turned off for access, following an access, they are turned back on and the Tevatron is ramped and brought through a ``dry squeeze'', where the low-beta quads are ramped without beam in the Tevatron.  The same process may be repeated with proton beam, referred to as a ``wet squeeze'', in order to check and correct orbits.  If corrections are needed, another wet squeeze is performed.  Figure~\ref{wetsqueeze} shows that the initial luminosity of a store following an access is generally about 3\% higher if a wet squeeze is performed.  Whether or not this slight increase in initial luminosity is worth the time it takes to perform the wet squeeze(s), the fact that going through the process makes it less likely to develop problems during shot setup makes it worth the effort.

\section{Results}

\begin{figure*}[h]
\includegraphics[width=1.1\linewidth]{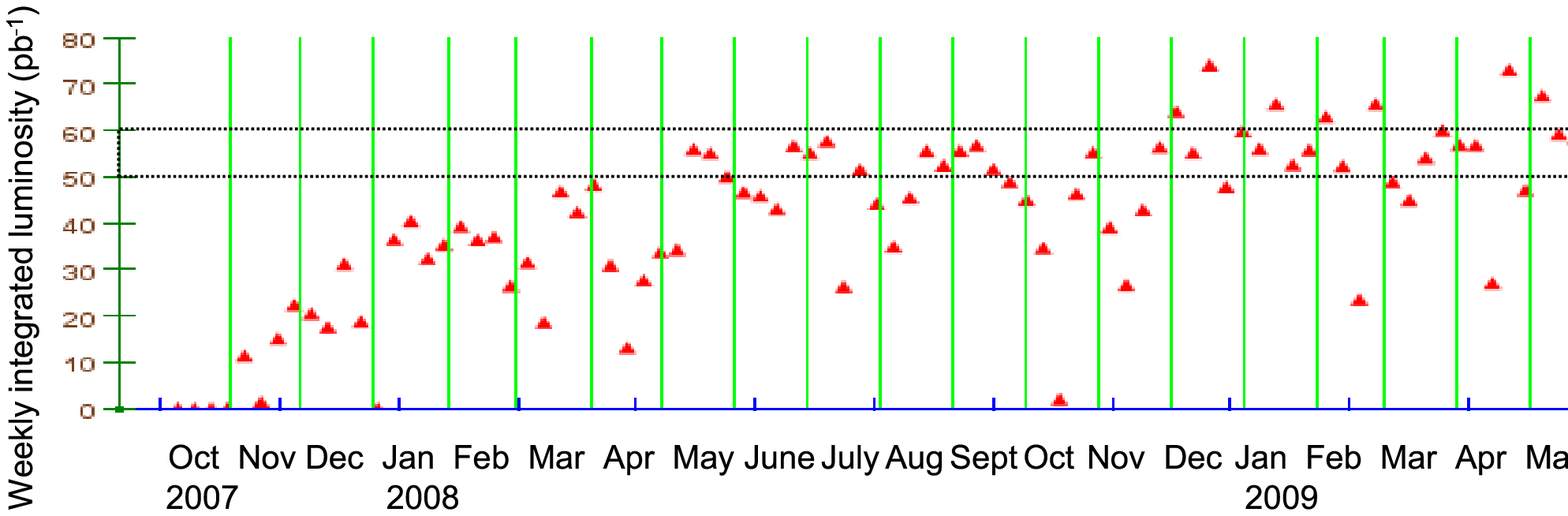}
\vglue -2.5in
\caption{Weekly integrated luminosity over the period from the 2007 shutdown to the 2009 shutdown.} \label{weeklylum}
\end{figure*}

Figure~\ref{weeklylum} Shows the weekly integrated luminosity from Oct 2007 to present.  Three distinct periods can be observed: first where luminosities were below 50 pb$^{-1}/$wk, second where the highest integrated luminosities we in the range 50-60 pb$^{-1}/$wk, and finally where 50-60 pb$^{-1}/$wk is our average.  The second period begins when we implemented the model described in Sec.~\ref{model} around the end of April 2008 (note that the machine uptime early in that period was low, less than 100 hrs/wk through the second week in May).  The third period begins after the shutdown in October 2008 when the aperture restriction near CDF was removed.  Most of the other improvements mentioned above were also implemented during this third period.

\begin{figure}[t]
\includegraphics[width=1.3\linewidth]{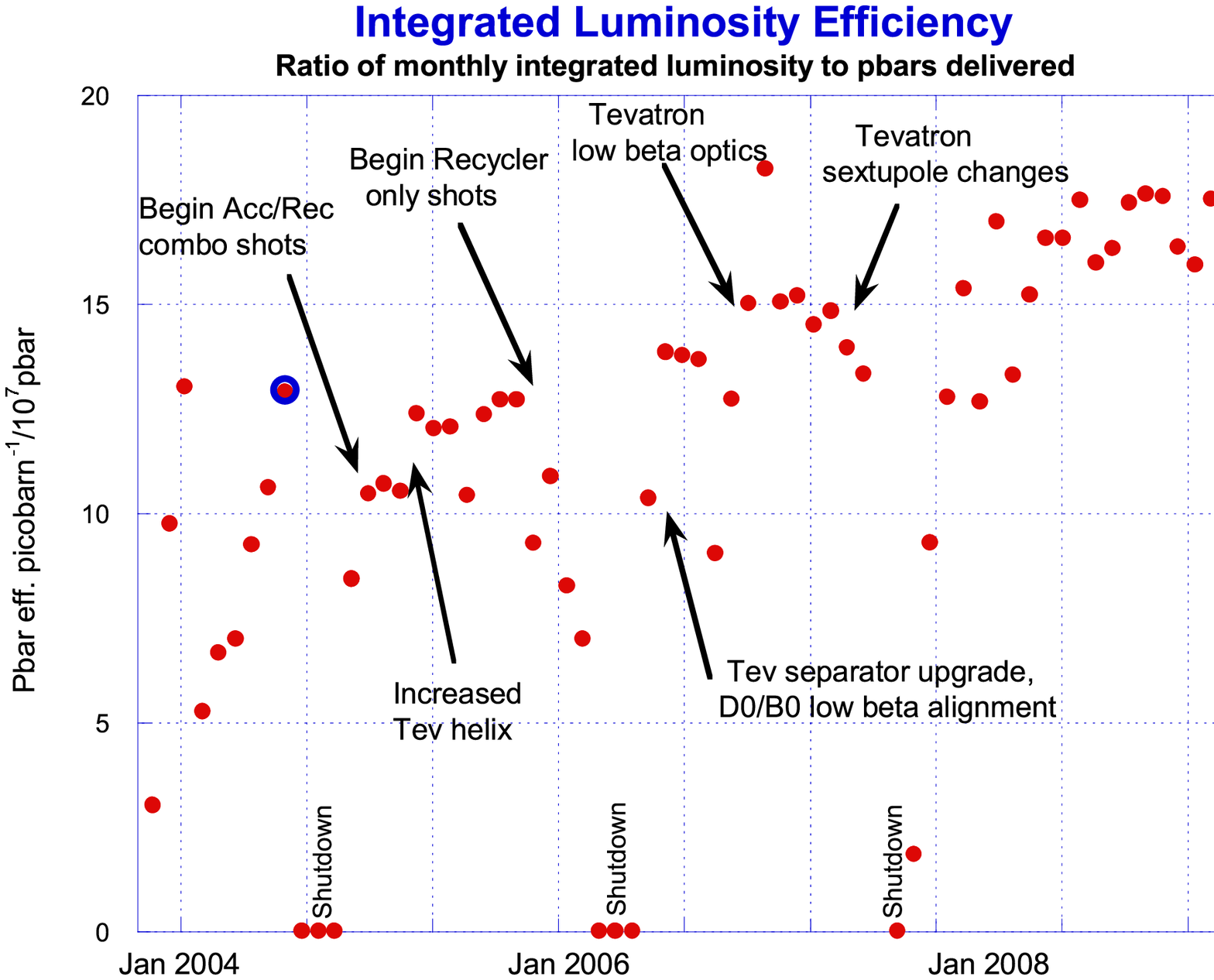}
\vglue -0.3in
\caption{Ratio of monthly integrated luminosity to antiprotons delivered to the Recycler.} \label{jimplot}
\end{figure}

Figure~\ref{jimplot} shows the effective use of available antiprotons in terms of integrated luminosity.  Looking at the period of April 2008 and later, we see that antiprotons are being used more efficiently since the model has been employed.

\section{Conclusions}
We have a model for optimizing integrated luminosity at the Fermilab Tevatron which is used to determine the target number of antiprotons for terminating a collider store and putting in a new one.  The implementation of this model has led to an improvement of approximately 35\% in integrated luminosity.

Operational changes have increased overall antiproton production, including optimizing pbar transfers, speeding up the time needed to prepare for and execute a transfer, and a new method for leaving behind a fraction of the antiprotons in the Recycler when extracting for a Tevatron store to allow more antiprotons in the Recycler than we want to use for the collider shot.

Other recent operational improvements include decreasing collider shot setup time, reducing beam-beam effects by making the proton and antiproton brightnesses more compatible, e.g., by scraping the proton beam to smaller emittance, as well as efforts towards consistency and reliability.

\begin{figure}[t]
\includegraphics[width=\linewidth]{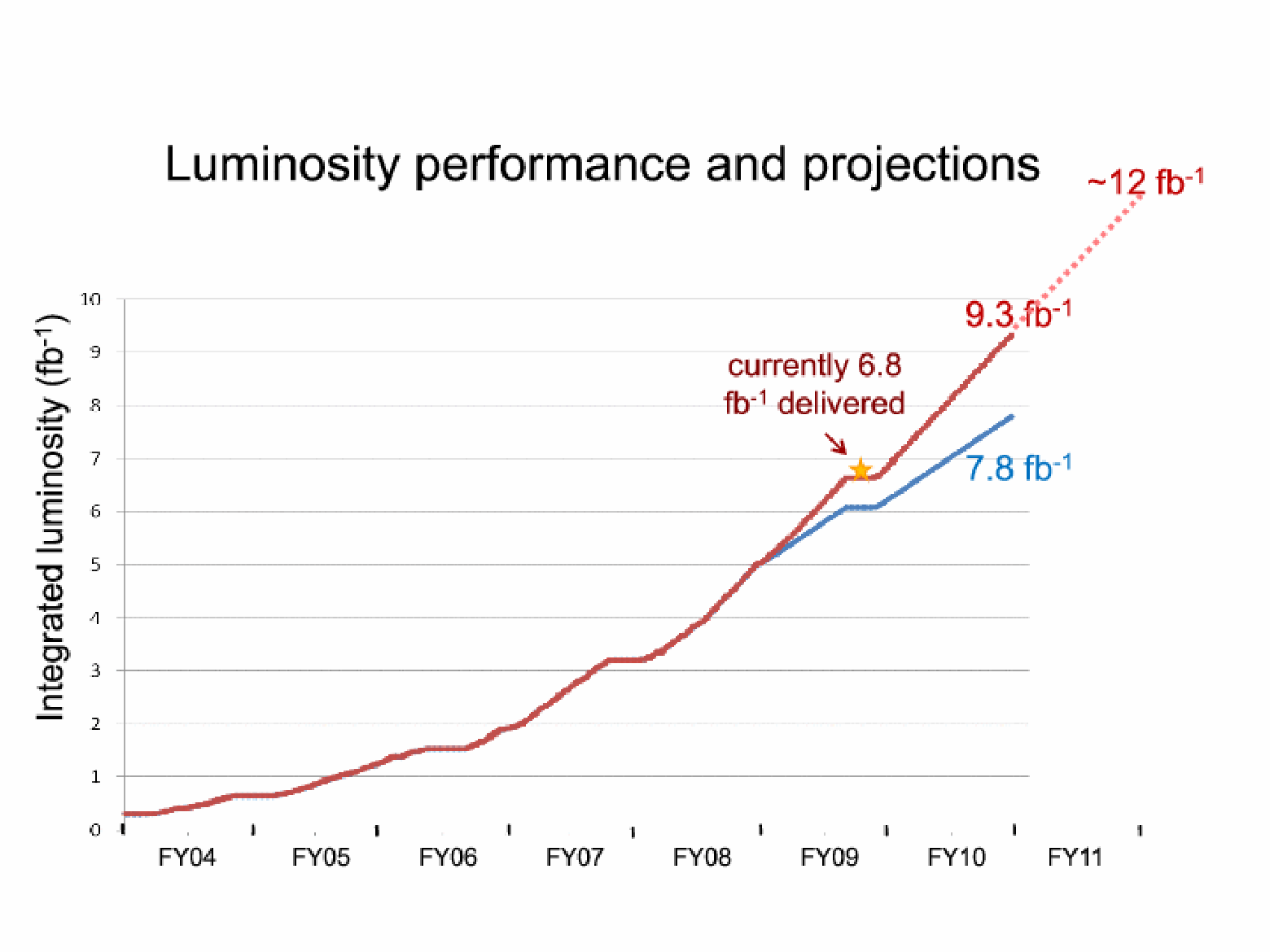}
\vglue -0.2in
\caption{Luminosity performance and projections.} \label{lumprojection}
\end{figure}

Figure~\ref{lumprojection} shows that if the Tevatron continues running straight through FY2011, we are on track to reach about 12 fb$^{-1}$ even with no further improvements.  However, we continue to work on operational improvements similar to what has been shown here and hope to continue to surpass expectations.

\begin{acknowledgments}
Many thanks to my partner Run Coordinator, Cons Gattuso, who had the inspiration for many of these improvements and many more to come.
\end{acknowledgments}

\bigskip 

\end{document}